\documentstyle[pre,aps,twocolumn]{revtex}
\input epsf

\begin{document}

\title{Passive scalar intermittency in low temperature helium flows}

\author{F. Moisy, H. Willaime, J.S. Andersen$^{*}$ and P. Tabeling}

\address{Laboratoire de Physique Statistique, \'Ecole Normale 
Sup\'erieure, 24 rue Lhomond, 75231 Paris Cedex 05 (France),\\
$^*$Centre for Chaos and Turbulence Studies, Niels Bohr Institute, 
Blegdamsvej 17-19, DK-2100 K{\o}benhavn {\O} (Denmark).}

\date{\today}
\maketitle

\begin{abstract}
We report new measurements of turbulent mixing of temperature
fluctuations in a low temperature helium gas experiment, spanning a
range of microscale Reynolds number, $R_{\lambda}$, from 100 to 650. 
The exponents $\xi_{n}$ of the temperature structure functions
$\langle |\theta(x+r)-\theta(x)|^n \rangle \sim r^{\xi_{n}}$ are shown
to saturate to $\xi_{\infty} \simeq 1.45 \pm 0.1$ for the highest
orders, $n \sim 10$.  This saturation is a signature of statistics
dominated by front-like structures, the cliffs.  Statistics of the
cliff characteristics are performed, particularly their width are
shown to scale as the Kolmogorov length scale.
\end{abstract}

\narrowtext

The strong intermittency of a passive scalar field advected by
a turbulent flow has recently received considerable
attention~\cite{Shraiman99,Warhaft2000}.  Two facets of this
intermittency are the persistence of small scale
anisotropy~\cite{Mestayer76,Gibson77} and the anomalous scaling of the
structure functions \mbox{$\langle |\theta(x+r)-\theta(x)|^n
\rangle$}~\cite{Antonia84,Kraichnan94}.  It is well established that
the persistence of scalar gradient skewness arises from the
ramp-and-cliff structures~\cite{Mestayer76,Gibson77}, {\it i.e} high
scalar jumps separated by well mixed regions.  The observed
preferential alignment of cliffs with the large scale
gradient~\cite{Holzer94,Pumir94} apparently prevents a universal
description of the odd order statistics, reflecting the asymetry of
scalar fluctuations.  However, the genericity of cliffs in ``scalar
turbulence'' raises the issue of their influence on high-order scaling
of even moments, and of the possible universality of the suspected
saturation of high order exponents~\cite{Yakhot97,Celani2000,caro}. 
Laboratory experiments able to cover a wide range of Reynolds numbers
in well controlled conditions appear to be crucial to address this
point.  We report in this Letter new measurements of turbulent mixing
of a scalar field, namely temperature fluctuations, performed in a low
temperature helium gas experiment.  Such measurements of temperature
fluctuations, as a passively advected scalar field, are performed for
the first time in low temperature helium gas~\cite{DonnellySR},
opening new and encouraging perspectives in investigation of turbulent
mixing in high Reynolds number flows.

\begin{figure} [b]
\epsfxsize=8cm 
\centerline{ \epsffile{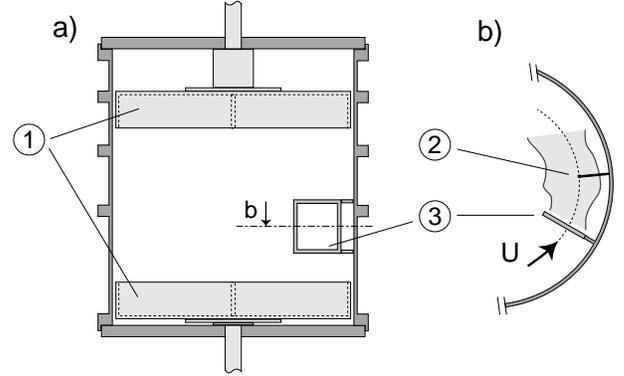}}
\vspace{3mm}
\caption{(a)~Sketch of the experiment.  (b)~Upper view, in
corotating mode.  (1)~Disks, (2)~thermometer, (3)~heated grid.}
\label{fig:dessinmanip}
\end{figure}

The set-up we use is the same as the one described in
Ref.~\cite{Zocchi94}, with an additional apparatus to induce
temperature fluctuations.  The flow takes place in a cylindrical
vessel and is driven by two rotating disks, as sketched in
Fig.~\ref{fig:dessinmanip}.  The disks, 20~cm in diameter and spaced
13.1~cm apart, are mounted with 6 radial blades.  The cell is filled
with helium gas, held at a controlled pressure.  Temperature
measurements are performed with a specially designed cold-wire
thermometer in the thermal wake of a heated grid.  The grid is made of
a double nichrome wire net, stretched across a $4.0 \times 4.5$~cm
frame.  The wire diameter is 250~$\mu$m, and the mesh size, $M$, is
2.0~mm.

The thermometer is located 20 mesh sizes downstream from the grid, at
a distance of 2.2~cm from the wall.  It has been designed as the
velocity probes described in Ref.~\cite{Zocchi94}.  It consists in a
thin carbon fiber, 7~$\mu$m in diameter, stretched across a 2~mm rigid
frame.  A metallic deposition of gold and silver of about
2000~\AA~thick covers the whole fiber, except for a 7~$\mu$m long
central area.  Temperature fluctuations are measured from resistance
fluctuations, by the way of a constant current Wheatstone bridge.  The
tuning of the current has received considerable attention, in order to
ensure an high enough signal, without suffering from velocity
contamination.  In practice, the optimal value of the current is
chosen in order to maximize the flatness factor of the temperature
derivative.  Temperature resolution is estimated, from the
high-frequency white noise level, to around 100~$\mu$K.
The spatial resolution is limited by the probe size, which
is at least twice smaller than the Kolmogorov scale in the present
experiments.  The frequency response is limited by the constant
current amplifier to 10~kHz, a value comfortably high to resolve the
highest frequency of the temperature fluctuations.

Since the same probe, with different electronic devices, is used as
both thermometer and anemometer, we are able to perform temperature
and velocity measurements at the same point in the same flow
conditions.  We have measured the various turbulence characteristics,
such as the microscale Reynolds number $R_{\lambda}$ and the
Kolmogorov scale $\eta$, for the same disks rotation rates and
kinematic viscosity $\nu$, allowing us to characterize each
temperature data set.  We define here $R_{\lambda} = u' \lambda / \nu$
the Reynolds number based on the Taylor scale $\lambda=u'(\nu /
\epsilon)^{1/2}$, where $u'$ is the rms of the velocity fluctuations,
$\epsilon$ the mean dissipation rate and $\eta=(\nu^{3}/\epsilon)^{1/4}$.

The disks rotation can be set in two modes, co- and counter-rotating,
hereafter denoted by COR and CTR. In the corotating mode, the two
disks are rotating in the same direction at the same speed, so that
the flow can be thought as a solid-like rotation.  The fluctuation
rate $u' / \langle U \rangle$ is found to lie around 10~\% in this
case, a value close to the one obtained in a previous experiment in a
similar set-up~\cite{Willaime99}.  In the counter-rotating mode, the
turbulence intensity is much higher, but the speed ratio is tuned so
that a strong mean advection remains at the height of the grid and the
thermometer.  This ensures a low enough fluctuation rate, between 13
and 25~\%, allowing use of the Taylor hypothesis to convert temporal
fluctuations into spatial ones.  We have checked that the fluctuation
rate is the same with and without the grid, meaning that the grid
contribution to the turbulence can be considered as negligible.  The
only influence of the grid is injection of heated sheets in an already
turbulent flow.  Thus thermal turbulence originates in the mixing of
these hot layers behind each wire, over a typical distance of $M
\langle U \rangle /u' = $1--2~cm downstream.

The grid temperature has to be high enough to induce thermal
fluctuations giving rise to an acceptable signal to noise ratio of at
least 50~dB. The noise level of 100~$\mu$K requires a rms signal of
typically $\theta' = \langle \theta^2 \rangle^{1/2} \simeq 40$~mK at
the probe location, which is around 4~\% of the grid overheat.  Since
we are working with a closed flow, the heat transfer from the cell to
the cryostat has to be balanced by the power supplied to the grid, in
order to ensure thermal stationarity.  This constraint limits the
maximum duration of an experiment (up to 5 hours of continuous run),
and the range of velocity and fluid density; in practice, values of
$R_{\lambda}$ in the range 100--300 in the corotating case, and
200--650 in the counter-rotating case, are obtained.  Although much
lower than the highest $R_{\lambda}$ achievable in our set-up (up to
5000~\cite{Zocchi94}), we hope that future noise improvement of the
constant current amplifier will increase the upper bound well beyond
650.

\begin{table} [b]
\begin{tabular}{llllll}
File no.& $10^4 \times \kappa$ & $R_{\lambda}$ & $\langle U \rangle $ &
$\theta'$ & $10^{-6} \,N^{*}$\\
 & (cm$^{2}$/s) & & (cm/s) & (mK) & \\
\hline 
 1 COR & 75 & 105 & 16.3 & 45.5 & 0.84 \\ 
 2 COR & 17 & 280 & 27.4 & 56.8 & 14.7 \\ 
 3 CTR & 17 & 360 & 28.2 & 96.0 & 30.6 \\ 
 4 CTR & 17 & 650 & 34.4 & 69.0 & 85.7   
\end{tabular}
\caption{Typical experimental parameters.  $\kappa = \nu / \mbox{Pr}$
is the thermal diffusivity, $\nu$ the kinematic viscosity and
$\mbox{Pr} \simeq 0.8$.  The sample size $N^*$ is expressed as $N
\langle U \rangle / (2 \pi f_{s} \eta)$, where $N$ is the number of
data points of the sample and $f_{s}$ the sampling rate.}
\label{tab:tempfiles}
\end{table}

The temperature signal is filtered, then sampled and recorded on a 16~bit
acquisition board (ITC-18 from InstruTech), at a sampling rate between
1 and 10~kHz, a value at least twice the low-pass filter frequency. 
Special attention has been paid to the low noise level of the whole
amplifying and sampling channel.  In order to achieve correct
convergence of higher-order statistics, the sample sizes are of the
order of $10^7 - 10^8$ Kolmogorov time scales long. 
Table~\ref{tab:tempfiles} summarizes the characteristics of the
typical data sets used here.

In order to characterize the large scale of the temperature and
velocity fluctuations, quantities of interest are the integral scales
$\Lambda_{\theta}$ and $\Lambda_u$, defined from the autocorrelation
functions of temperature $\theta$ and longitudinal velocity $u$.  For
our whole data set, we measure $\Lambda_{\theta} = 7.0 \pm 1.5$~mm and
$\Lambda_u = 9.6 \pm 1.0$~mm, with no noticeable $R_{\lambda}$
dependence~\cite{Willaime99}.  This two values are rather close,
indicating that kinetic energy and temperature variance are injected
at the same large scale.  The Prandtl number of the gas being close to
unity, we are on the case where velocity and scalar fluctuations take
place on the same range of scales.

\begin{figure} [b]
\epsfxsize=8.6cm 
\centerline{ \epsffile{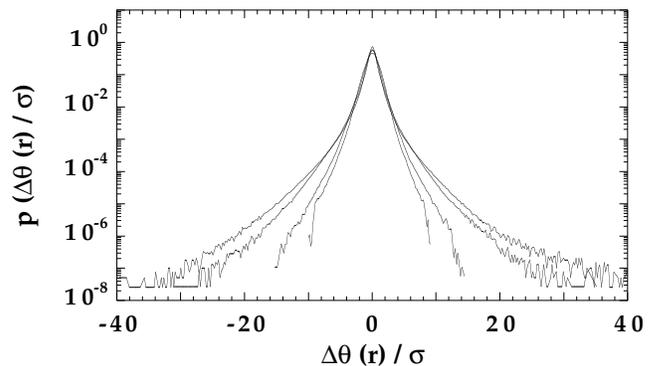}} 
\caption{Probability density functions of the normalized temperature
increments, for $R_{\lambda}=650$ (file \#4 of
Tab.~\ref{tab:tempfiles}).  From the inner to the outer pdf, $r/\eta =
10^4$, 600, 30 and 3.}
\label{fig:pdf141}
\end{figure}

The small scale intermittency of temperature fluctuations can be
characterized through their distributions at different scales. 
Figure~\ref{fig:pdf141} shows the probability density functions (pdf)
of temperature increments $\Delta \theta(r) = \theta(x+r)-\theta(x)$
for 4 different separations $r$, ranging from $r/\eta$ = 10$^4$ (large
scale), 600, 30 (typical inertial scales) down to 3 (close to the
dissipative scale).  Each distribution has been normalized by its
standard deviation $\sigma = \langle \Delta \theta(r)^2
\rangle^{1/2}$.  These pdf are strongly not self-similar, reflecting
strong intermittency effects.  The width of the tails is remarkable,
showing scalar jumps of amplitude up to 40 times the standard
deviation at the smallest scale.  We can note the slight asymmetry of
the distributions, linked to the well known property of small scale
persistence of anisotropy~\cite{Warhaft2000}.

\begin{figure} [b]
\epsfxsize=7cm 
\centerline{ \epsffile{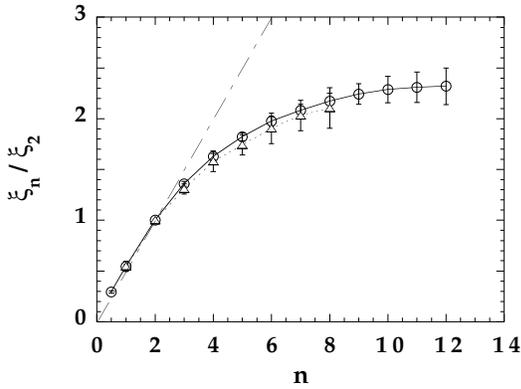}} 
\caption{Temperature structure function exponents, $\xi_{n}$,
normalized by $\xi_{2}$.  $\bigcirc$:~$R_{\lambda}=280$ in
COR mode, $\triangle$:~$R_{\lambda}=650$ in CTR mode.  The dashed line
indicates the Corrsin-Obukhov scaling $n/2$.}
\label{fig:xnx2}
\end{figure}

\begin{figure} [b]
\epsfxsize=8.6cm 
\centerline{ \epsffile{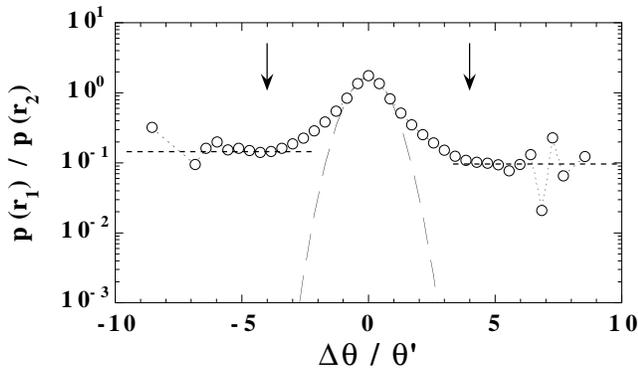}} 
\caption{Ratio of two pdfs, at different inertial scales, $r_{1}=45 \eta$
and $r_{2}=246 \eta$ ($R_{\lambda}=280$).  The long dashed curve is a
gaussian fit, and the horizontal dashed lines indicate the constant
ratio of the pdf tails.  Vertical arrows, at $\Delta \theta \simeq 4
\theta'$, indicate the dominant contribution to the 8th order moment.}
\label{fig:pdfratio}
\end{figure}

The evolution of these pdfs is characterized by the structure
functions, defined as \mbox{$S_{n} (r) = \langle |\theta(x+r)-\theta(x)|^n
\rangle$}, where angular brackets denote space (time) average.  They
are found to follow power law in terms of the scale $r$, namely
$S_{n} (r) \sim r^{\xi_{n}}$.  The measured structure
function exponents $\xi_{n}$, divided by $\xi_2$, are shown in
Fig.~\ref{fig:xnx2}, in the COR ($R_{\lambda}=280$) and CTR
($R_{\lambda}=650$) cases.  The exponents are defined by plotting the
compensated structure functions $r^{-\xi_{n}} S_{n} (r)$ and
tuning the value of the exponent to obtain a well defined plateau for
inertial separations.  This procedure allows to estimate the error bar
for each order.  Although the second order exponent presents some
scatter (with a systematic increase from 0.45 to 0.65 with increasing
$R_{\lambda}$), the normalization of the higher order exponents by
$\xi_{2}$ provides an excellent collapse for the different
$R_{\lambda}$.  It is remarkable that, for comparable sample sizes,
the highest available order is much lower at higher Reynolds number:
the fluctuations becomes much more intermittent and the convergence
becomes poorer.  For $R_{\lambda}=650$ we have to restrict to $n \leq
8$, whereas $n=12$ can be achieved for $R_{\lambda}=280$.  The two
sets of exponents are consistent within error bars for $n \leq 8$,
meaning that the large scale properties of the two flow configurations
do not affect these inertial range statistics.

The values of $\xi_n / \xi_{2}$ are found to strongly depart from the
linear law $n/2$, and the gap increases with the order, a usual
signature of inertial range intermittency of the passive
scalar\cite{Warhaft2000}.  Furthermore, the exponents are found
to increase extremely slowly with the order, strongly suggesting a
saturation, for $n$ around 10, at a value
\begin{equation}
\xi_{\infty} = (2.3 \pm 0.1) \xi_{2} \simeq 1.45 \pm 0.1
\label{eq:xiinf}
\end{equation}
(where $\xi_2$ is considered equal to 2/3).  This observation only
relies on the COR data set, but we have noted that no deviation
appears between the COR and CTR data for $n \leq 8$.  A saturation of
the structure function exponents is a signature of statistics
dominated by shock-like structures, the thermal cliffs.  A consequence
of this saturation is a constant ratio of the far tails of the pdfs
for inertial range separations.  The ratio of two pdfs, for
separations $r_{1}$ and $r_{2}$ well into the inertial range, is
displayed in Fig.~\ref{fig:pdfratio}.  For temperature increments
$|\Delta \theta| > 4 \theta'$, this ratio tends towards a constant. 
It must be noted that the $|\Delta \theta| \simeq 4 \theta'$ parts of
this distribution give the dominant contribution to the 8th order
moment (see the vertical arrows), {\it i.e} they maximize the
integrant $|\Delta \theta|^8 p(\Delta \theta)$.  This additional test
confirms the observed saturation of high order exponents.

\begin{figure} [b] \epsfxsize=8.6cm
\centerline{\epsffile{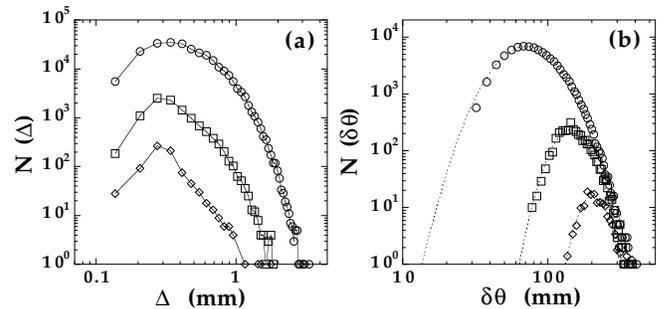}}
\caption{Histograms of cliff width (a) and amplitude (b) for values of
the threshold $s=3$ ($\bigcirc$), 7 ($\Box$) and 11 ($\diamond$).  The
dashed curves on the figure (b) are log-normal fits.}
\label{fig:pdf_cliffs}
\end{figure}

The observed trend towards a saturation of high order exponents
motivates a detailed analysis of the cliffs.  We have performed
statistics of the cliff characteristics, defined as the strongest
temperature gradients, singled out from time series of temperature
fluctuations.  We define the cliffs from a simple threshold on the
temperature derivative,
\begin{equation}
|\partial \theta / \partial x | > s \, \langle (\partial \theta / 
\partial x)^{2} \rangle^{1/2},
\label{eq:thres}
\end{equation}
where $s$ is a non dimensional constant.  Spatial derivatives are
obtained from temporal ones using the Taylor hypothesis, and the
gradients are estimated from finite difference over the smallest
resolved separation.  We define the cliff amplitude $\delta \theta$ as
the difference between the two extrema surrounding the gradient
satisfying~(\ref{eq:thres}), and the cliff width $\Delta$ such that
the temperature derivative takes values exceeding 90~\% of its local
maximum.  Histograms of amplitudes and widths are shown in
Fig.~\ref{fig:pdf_cliffs}, for three values of the threshold $s$, at a
Reynolds number $R_{\lambda} = 280$.

We first note that both amplitude and width histograms can be well
fitted by log-normal distributions (given by a parabola in log-log
coordinates).  The mean amplitude of the cliffs is of order $\theta'$,
and is found to increase with the threshold $s$.  On the other hand,
the cliff width remains constant for different thresholds. This means
that the strength of the gradients depends mainly on the amplitude of
the scalar jump, and not on its width.

\begin{figure} [b]
\epsfxsize=8.6cm 
\centerline{ \epsffile{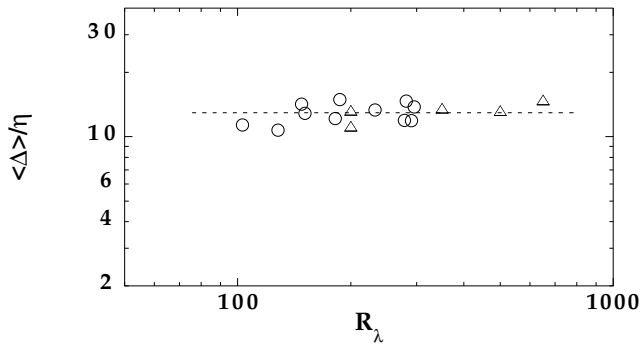}} 
\vspace{2mm}
\caption{Mean cliff width divided by the Kolmogorov length scale, as a 
function of $R_{\lambda}$. $\bigcirc$:~COR,
$\triangle$:~CTR.  The dashed line is the average, $\langle 
\Delta \rangle / \eta \simeq 13$.}
\label{fig:cliffwidth}
\end{figure}

A central issue concerning the cliffs is the Reynolds number
dependence of their width.  Figure~\ref{fig:cliffwidth} shows the mean
cliff width $\langle \Delta \rangle$, divided by the Kolmogorov scale
$\eta$, for $R_{\lambda}$ ranging from 100 to 650.  This plot extends a
preliminary study~\cite{moisyETC} performed only on the COR mode.  We can
see a well defined plateau,
\begin{equation}
\langle \Delta \rangle = (13 \pm 3) \eta.
\end{equation}
The scatter is moderate, and probably reflects the difficulty to resolve
properly the smallest scales.  However, this plot confirms that
the mean cliff width follows a $R_{\lambda}^{-3/2}$ scaling.  This law
holds for values of the threshold $s$ from 3 up to 40 in the case of
the highest $R_{\lambda}$ (see Fig.~\ref{fig:pdf141}).  We can note
that the data from the two configurations overlap on the central
region $R_{\lambda} = 200-300$, suggesting that this small scale
characteristics is not affected by the large scale properties of the
flow.

To summarize, we have performed for the first time measurements of
turbulent mixing of temperature, considered as a passive scalar field,
in a low temperature helium experiment.  The study of inertial range
statistics of the temperature increments gives evidence of a
saturation of the high order exponents, to a value $\xi_{\infty}
\simeq 1.45 \pm 0.1$.  This observation reveals that inertial range
statistics are dominated by the cliffs, concentrating large scalar
jumps over small distance~\cite{Yakhot97,Celani2000,caro}.  The cliff
widths are shown to scale as the Kolmogorov length scale, in the
whole range of observed $R_{\lambda}$ (100--650), suggesting that the
strongest cliffs remain concentrated over the smallest scale of the
flow (see also Ref.~\cite{Nieuwstadt}).  This observation may have
important consequences for processes such as reactive mixing or
combustion, where the reaction rate is enhanced where concentration
gradients are strong.  Further insight into the cliffs contribution to
the saturation of the high order structure function exponents needs a
detailed study of their morphology and spatial distribution. 
Preliminary results~\cite{moisyETC}, based on the waiting time between
cliffs from the temperature time series, strongly suggests
self-similar clustering for inertial separations.

The authors thank B.~Shraiman, V.~Hakim, M.~Vergassola, P.~Castiglione
and M.C.~Jullien for fruitful discussions.  This work has been
supported by Ecole Normale Sup\'erieure, CNRS, the Universities Paris
6 and Paris 7, and the European Commission's TMR programme, Contract
no.~ERBFMRXCT980175 ``Intermittency''.

\end{document}